\documentclass[serif]{beamer}
\usepackage{amsmath}
\usepackage{url}
\usepackage{caption}
\usepackage[T1]{fontenc}
\usepackage{physics}
\usepackage[style=verbose,backend=bibtex]{biblatex}
\makeatletter
\renewcommand{\@makefnmark}{\makebox{\normalfont[\@thefnmark]}}
\makeatother

\newtheorem{proposition}[theorem]{Proposition}

\mode<presentation> {

\usetheme{Madrid}

\setbeamertemplate{navigation symbols}{} 
}

\usepackage{graphicx} 
\usepackage{booktabs} 

\numberwithin{equation}{section}

\title[Symmetries and Riemann Invariants]{Symmetries and Riemann Invariants} 

\author{A. Michel Grundland} 

\institute[] 
{
In collaboration with Javier de Lucas (Department of Mathematical Methods in Physics, University of Warsaw, Poland)

\vspace{0.5cm}

Centre de Recherches Mathématiques, Université de Montréal \\
and \\
Département de Mathématiques et d'Informatique \\
Université du Québec à Trois-Rivières  \\
\vspace{0.5cm}

Conference dedicated to the memory of Pavel Winternitz. \\
Centre de Recherches Mathématiques, Université de Montréal, July 2021

}
\date{}

\begin{document}

\begin{frame}
\titlepage 
\end{frame}



\begin{frame}
\begin{center}
  \textit{\guillemotleft Where symmetries are to be found, \\ there is hope for discovery \guillemotright}  

\vspace{1cm}
\textit{\guillemotleft Là où les symétries peuvent êtres trouvées, \\ il y a un espoir de découverte  \guillemotright}
\end{center}

\flushright
Pavel Winternitz
\end{frame}


\begin{frame}
\frametitle{Objectives of the Study}

\begin{itemize}
\item To establish a relation between two approaches to the construction of Riemann k-wave solutions of hydrodynamic-type systems, namely the symmetry reduction method and the generalized method of characteristics. 
\end{itemize}
\end{frame}




\begin{frame}
\frametitle{Discussions with Pavel}

Much has already been said here about the life and work of Pavel Winternitz. I would only like to underline that Pavel’s impact has to be measured not only by his research results, but also by his contribution to the training of young scientists and the inspiration he has provided to them. His mentorship and friendship meant a lot to a lot of people, myself included. My first meeting with him, in Banff in August 1980, opened new doors for me and led to our later collaboration. It resulted in a series of our joint papers, co-authored also with other researchers like J. Harnad, J. Tuszynski, E. Infeld, W. Zakrewski, M. Sheftel and P. Tempesta. What was remarkable about Pavel was that he was always a source of interesting new research ideas, attracting other scientists. He had a gift for bringing people together. Our joint works in the period (1982-2003) was based on the approach developed by Pavel with his collaborators J. Patera, R. Sharp and H. Zassenhaus, namely the classification of Lie subalgebras by conjugacy classes and their use in the symmetry reduction procedure. Our topics included the analysis of 

\end{frame}


\begin{frame}
\frametitle{Discussions with Pavel}

nonlinear relativistically invariant equations, nonlinear  and  field models, Landau-Ginzburg equations, hydrodynamic-type systems and partially and conditionally invariant solutions. 
\\~\\
In this talk, I will focus on the subject inspired by my discussions with Pavel during our first meeting in 1980, a subject to which I have returned several times since then, always encouraged by Pavel’s interest and his questions. This subject concerns the connection between two seemingly unrelated concepts, namely the symmetry reduction method and the method of characteristics for hydrodynamic-type systems in many dimensions. I will present a problem on multiple Riemann waves in view of these two approaches.
I will add that Pavel loved not only mathematics, but also nature – both sea and mountains, and this investigation would probably never have happened had it not been for our discussions during the wonderful two-days hike in the Rocky Mountains that we took that summer, together with Michael and Peter.

\end{frame}



\section{Multiple Riemann Waves for Hydrodynamic Systems}
\begin{frame}
\frametitle{1. Multiple Riemann Waves for Hydrodynamic Systems}
Riemann waves represent a very important class of solutions of nonlinear 1st-order hyperbolic systems of PDEs 
\begin{align} \label{1.1}
A^i(u)u_i=0,  \qquad i=1,...,p,
\end{align}
\begin{align*}
&\text{where} \quad  u_i=\pdv{u}{x^i}, \quad A^1,...,A^p \, -\, q \times q \, \text{matrix functions of } u, \\
&x=(x^1,...,x^p)\in E \subset \mathbb{R}^p, \qquad u=(u^1,...,u^q)\in \mathcal{H} \subset \mathbb{R}^q.
\end{align*}
Here we use the summation convention.
For a homogeneous system \eqref{1.1}, a \textbf{Riemann simple wave} solution is defined by the equation
\begin{align} \label{1.2}
u=f(r(x,u)),
\end{align}
where $f:\mathbb{R}\rightarrow \mathbb{R}^q$ and the function $r: \mathbb{R}^{p+q} \rightarrow \mathbb{R}$ given by
\begin{align} \label{1.3}
r(x,u)= \lambda_i (u) x^i,
\end{align}

\end{frame}


\begin{frame}
\frametitle{1. Multiple Riemann Waves for Hydrodynamic Systems}
is called a \textbf{Riemann invariant} of the wave vector $\lambda=(\lambda_1,...,\lambda_p)$ such that
\vspace{-0.2cm}
\begin{align} \label{1.4}
\ker \left( \lambda_i A^i(u) \right) \neq 0.
\end{align}
These solutions have rank at most equal to one : $\pdv{u}{x^i} \sim \lambda_i \dv{f^\alpha}{r}$.

The problem of superposition of simple Riemann waves was first solved through the generalized method of characteristics. 
This method relies on treating Riemann invariants as new dependent variables which remain constant along appropriate characteristic curves of the basic system \eqref{1.1}. This leads to the reduction of the dimensionality of the problem. 
The construction of Riemann $k$-wave solutions can be summarized as follows :
\vspace{-0.2cm} 
\begin{enumerate}
\item[\textbf{1)}] We assume that there exist $kp$ functions $\lambda^s_i: \mathfrak{B}\subset \mathcal{H} \rightarrow \mathbb{R}$ and $kq$ functions $\gamma^\alpha_s: \mathfrak{B}\subset \mathcal{H} \rightarrow \mathbb{R}$ such that the \textbf{wave relation}
\begin{align} \label{1.5}
\left(A^{i\beta}_\alpha(u) \lambda^s_i \right)\gamma^\alpha_{(s)} =0, \quad \gamma_s=(\gamma^1_s,...,\gamma^q_s), \quad \lambda^s=(\lambda^s_1,...,\lambda^s_p).
\end{align}
$s=1,...,k, \quad \alpha, \beta=1,...,q,\quad$ holds.
\end{enumerate}
\end{frame}


\begin{frame}
\frametitle{1. Multiple Riemann Waves for Hydrodynamic Systems}
\begin{enumerate}
\item[\textbf{2)}] Assuming that we have $k$ linearly independent functions $\lambda^s$ and $\gamma_s$ and identifying the vectors $\gamma_s$ with vector fields on the hodograph space $\mathfrak{B}\subset \mathbb{R}^q$, we verify whether the involutivity conditions for the tangent map 
\begin{align} \label{1.6}
du^\alpha(x)=\sum^k_{s=1} \xi^s(x) \gamma^\alpha_s (u) \lambda^s_\mu (u) dx^\mu,
\end{align}
are satisfied 
\begin{align} \label{1.7}
[ \gamma_s, \gamma_r]\in  \text{span}\left\lbrace \gamma_s, \gamma_r \right\rbrace, \quad \lambda^s_{,\gamma_r} \in\text{span} \left\lbrace \lambda^s, \lambda^r \right\rbrace, \quad r\neq s=1,...,k.
\end{align}
\item[\textbf{3)}] We choose a holonomic system for the vector fields $\left\lbrace\gamma_1, ..., \gamma_k \right\rbrace$ by requiring a proper length for each vector $\gamma_s$ such that $[ \gamma_s, \gamma_r]=0, \, s \neq r=1,...,k$. 
This implies that there exists a manifold $S$ which can be explicitly parametrized by solving the PDEs.  
\end{enumerate}
\end{frame}


\begin{frame}
\frametitle{1. Multiple Riemann Waves for Hydrodynamic Systems}
\begin{align} \label{1.8}
\pdv{f^\alpha}{r^s}=\gamma^\alpha_s (f^1,..., f^k),
\end{align} 
whose solution is defined by the equation
\begin{align} \label{1.9}
u=\left(f^1(r^1,...,r^k),...,f^q(r^1,...,r^k)\right).
\end{align}
\begin{enumerate}
\item[\textbf{4)}] Next, we look for the most general solution of the 2nd-order linear system of PDEs (no summation) for the wave vectors $\lambda^s$ (a consequence of \eqref{1.7})
\begin{align} \label{1.10}
\pdv[2]{\lambda^s}{r^s}{r^\ell}+\alpha^s_\ell \pdv{\lambda^s}{r^{s}}+\beta^s_\ell \pdv{\lambda^s}{r^\ell}=0, \qquad  s\neq \ell=1,...,k,
\end{align}
where $\alpha^s_\ell$ and $\beta^s_\ell$ are functions of $r^1,...,r^k$. 
\item[\textbf{5)}] The general solution for the $\lambda^s$ of \eqref{1.10} allows us in turn to determine the solution of the Pfaffian system 
\vspace{-0.2cm}
\begin{align} \label{1.11}
dr^s=\xi^s(x)\lambda^s(r^1,...,r^k), \qquad s=1,...,k,
\vspace{-0.2cm}
\end{align}
\end{enumerate}
\end{frame}


\begin{frame}
\frametitle{1. Multiple Riemann Waves for Hydrodynamic Systems}
which has the implicit form 
\begin{align}\label{1.12}
\lambda^s_\mu(r^1,...,r^k)x^\mu=\psi^s(r^1,...,r^k),
\end{align}
where the $\psi^s$ are arbitrary functions of $r^1,...,r^k$.
The quantities $r^1,...,r^k$ constitute Riemann invariants as they are implicitly defined as functions of $x^1,...,x^p$ by \eqref{1.12} in the $E\subset \mathbb{R}^p$ space. \\~\\

Thus the construction of the \textbf{Riemann k-wave solution} is determined by the equation
\begin{align*}
u=f\left(r^1(x,u),...,r^k(x,u)\right),
\end{align*}
and the implicit relation \eqref{1.12}, where $\lambda^s$ satisfy the differential constraints \eqref{1.10}. This approach has produced many new analytic solutions of hydrodynamic models (e.g. equations in continuous media and nonlinear field equations).
\end{frame}


\section{Symmetries and Riemann Wave Solutions}
\begin{frame}
\frametitle{2. Symmetries and Riemann Wave Solutions} 
Now there arises the question of what additional results, if any, can be achieved by applying the symmetry group analysis to the problem of the construction of Riemann k-waves. \\

Let us fix a set of $k\leq p$ linearly independent wave vectors $\lambda^1,..., \lambda^k$, with corresponding Riemann invariants
\begin{align} \label{2.1}
r^1=\lambda^1_i(u) x^i, ..., r^k=\lambda^k_i(u) x^i, \qquad 1 \leq k \leq p.
\end{align}
The equation for a Riemann k-wave
\begin{align} \label{2.2}
u=f\left( r^1(x,u),...,r^k(x,u)\right),
\end{align}
defines a unique function $u(x)$ on a neighbourhood of $x=0$, for any function $f: \mathbb{R}^k\rightarrow \mathbb{R}^q$ and the tangent map is
\begin{align} 
\pdv{u^\alpha}{x^i}\,(x)=\left( \phi(x)^{-1}\right)^l_j \lambda^j_i (u(x)) \pdv{f^\alpha}{r^l}\,(r(x,u(x))),\label{2.3} \\
\phi(x)^i_j=\delta^i_j-\pdv{r^i}{u^\alpha}\,(x,u(x)) \pdv{f^\alpha}{r^j}\,(r(x,u(x))). \label{2.4}
\end{align}
\end{frame}


\begin{frame}
\frametitle{2. Symmetries and Riemann Wave Solutions}
Note that the rank of the solution $u(x)$ is at most equal to $k$. 
If the $(p-k)$ vector fields
\begin{align} \label{2.5}
\xi_a(u)=\left(\xi^1_a (u), ..., \xi^p_a (u) \right)^\intercal, \qquad a=1,...,p-k,
\end{align}
satisfy $\lambda^j_i\xi^i_a=0, \quad j=1,...,k$, then from the tangent map \eqref{2.3} we obtain
\begin{align}\label{2.7}
\xi^i_a(u(x)) \pdv{u^\alpha}{x^i}\, (x)=0, \qquad a=1,...,p-k.
\end{align}
This means that the graph $\left\lbrace(x,u(x))\right\rbrace$ is invariant under the $(p-k)$ vector fields
\begin{align}\label{2.8}
X_a=\xi^i_a(u) \pdv{}{x^i},  \qquad \text{on} \; \mathbb{R}^p\times \mathbb{R}^q , \quad a=1,...,p-k.
\end{align}
\end{frame}


\begin{frame}
\frametitle{2. Symmetries and Riemann Wave Solutions}

If $u(x)$ is the solution of the equation for a Riemann k-wave \eqref{2.2}, then the basic system \eqref{1.1} reduces to 
\begin{align} \label{2.9}
&A^i(u(x))\pdv{u^\alpha}{x^i}\, (x)=\left( \phi(x)^{-1}\right)^\ell_j \lambda^j_i (u(x)) A^i(u(x)) \pdv{f^\alpha}{r^\ell}\,(r(x,u(x)))=0, \\
&\phi(x)^i_j=\delta^i_j-\pdv{r^i}{u^\alpha}\,(x,u(x)) \pdv{f^\alpha}{r^j}\,(r(x,u(x))). \nonumber
\end{align}
Let us consider a coordinate transformation. 
Let us assume that the $k\times k$ matrix built in terms of the wave vectors $\Lambda=(\lambda^j_\ell)_{1 \leq \ell, j\leq k \leq p}$ is invertible.
\end{frame}


\begin{frame}
\frametitle{2. Symmetries and Riemann Wave Solutions}
The independent vector fields
\begin{align} \label{2.11}
X_a=\pdv{}{x^a}\, - \sum^k_{i,j=1} \left(\Lambda^{-1} \right)^\ell_j \lambda^j_a \pdv{}{x^\ell}, \qquad a=k+1,...,p,
\end{align}
have the form \eqref{2.8} with $\lambda^j_i \xi^i_a=0$, $j=1,...,k$. 
The functions
\begin{align} \label{2.12}
&\bar{x}^1=r^1(x,u),...,\bar{x}^k=r^k(x,u), \bar{x}^{k+1}=x^{k+1},...,\bar{x}^p=x^p, \\
&\bar{u}^1=u^1,...,\bar{u}^q=u^q, \nonumber
\end{align}
are coordinates on $\mathbb{R}^p \times \mathbb{R}^q$. This allows us to rectify the vector fields \eqref{2.11}
\begin{align} \label{2.13}
X_{k+1}=\pdv{}{\bar{x}^{k+1}}\, ,..., X_p=\pdv{}{\bar{x}^p}.
\end{align}
The $p$-dimensional manifold is invariant under $X_{k+1},...,X_p$ and is defined by the equations
\begin{align} \label{2.14}
\bar{u}=f(\bar{x}^1,..., \bar{x}^k ), \qquad \; \text{where} \; f: \mathbb{R}^k \rightarrow \mathbb{R}^q \; \text{is arbitrary}.
\end{align}

\end{frame}


\begin{frame}
\frametitle{2. Symmetries and Riemann Wave Solutions}
Thus the function $\bar{u}(\bar{x})$ is the general solution of the invariance conditions
\begin{align} \label{2.15}
\bar{u}_{k+1}=... =\bar{u}_p=0.
\end{align}
The initial system of PDEs \eqref{1.1} can be described in terms of the new coordinates $\bar{x}, \bar{u}$ by
\begin{align} \label{2.16}
\bar{A}^i(\bar{x}, \bar{u},\bar{u}_{\bar{x}})\bar{u}_i=0,
\end{align}
where
\begin{align}
&\bar{A}^s=\frac{Dr^s}{Dx^i}A^i, \qquad s=1,...,k \qquad \bar{A}^{k+1}=A^{k+1}, ..., \bar{A}^p=A^p, \nonumber \\
&\frac{Dr^i}{Dx^j}\Bigr|_{\bar{u}^{k+1}=...=\bar{u}^p=0}=(\phi^{-1})^i_\ell \lambda^\ell_j, \qquad \phi^i_j=\delta^i_j-\pdv{r^i}{u^\alpha} \bar{u}^\alpha_j. \nonumber
\end{align}
\end{frame}


\begin{frame}
\frametitle{2. Symmetries and Riemann Wave Solutions}
So, subjecting the system of PDE \eqref{2.16} to the invariant conditions \eqref{2.15} produces the overdetermined system of PDEs which defines Riemann k-wave solutions of the initial system \eqref{1.1}
\vspace{-0.4cm}
\begin{align*} \label{2.17}
&\sum^k_{i,j=1} \sum^p_{\ell=1} \left(\phi^{-1}\right)^i_j \lambda^j_\ell A^\ell \bar{u}_i=0,
\qquad \phi^i_j=\delta^i_j-\pdv{r^i}{u^\alpha} \bar{u}^\alpha_j,\\ 
&\bar{u}_{k+1}=...=\bar{u}_p=0.
\vspace{-0.8cm}
\end{align*}
\setcounter{footnote}{0} 
Numerous examples of new $k$-wave solutions for hydrodynamic-type systems in (3+1) dimensions have been obtained through this approach e.g. $\,$ \footcite{grundland2007conditional}.
This method has proved to have an advantage over the method of characteristics, since it results mainly from the relaxation of the 2nd-order differential constraints \eqref{1.10} for the wave functions $\lambda^s$. It delivers larger classes of solutions through this simpler procedure.

\end{frame}


\section{Conclusion}
\begin{frame}
\frametitle{Conclusion}
As we can see, this approach is a variant of the conditional symmetry method. 
I was encouraged in this investigation by Pavel who himself worked extensively on conditional symmetries for soliton equations. \\~\\

A natural continuation of this study is to extend it to non-autonomous hydrodynamic systems. This work was initiated in our joint work with Pavel and M. Sheftel \, \footcite{grundland2000invariant}. \\~\\

Currently, I am working on this topic with Javier de Lucas from the University of Warsaw. We have shown that the notion of Riemann invariants can be adapted to these types of systems and Riemann waves can be constructed for them via the appropriate version of the conditional symmetry method. 

\end{frame}


\begin{frame}
\frametitle{Conclusion}
At the end I have to say that it is very difficult for me to get used to the absence of Pavel, as it is for all of us here at the Centre de Recherches Mathématiques in Montreal where Pavel was an everyday presence. \\
We benefited from his support, his knowledge and advice in matters of science and life. 
We enjoyed his wit and his great sense of humor, conversations with him and his comments on all subjects - books, movies, event of the day and history. 
This place will not be the same without him.
But, as we have seen at this event, he will live in his students and all the scientists whom he inspired. \\~\\
Milada, thank you for everything you have done for Pavel and all of us, for your wonderful hospitality and your lovely spirit. 
My heart goes out to you, to Peter and Micheal and your whole family. 
\end{frame}


\begin{frame}
\frametitle{Conclusion}
\centering \textbf{Thank you for your attention.}
\end{frame}
\end{document}